%% file: 0-oag-bench.tex
\documentclass[sigconf]{acmart}

\AtBeginDocument{%
  \providecommand\BibTeX{{%
    \normalfont B\kern-0.5em{\scshape i\kern-0.25em b}\kern-0.8em\TeX}}}

\setcopyright{acmcopyright}

\acmConference[Conference acronym 'XX]{Make sure to enter the correct
  conference title from your rights confirmation emai}{June 03--05,
  2018}{Woodstock, NY}
\usepackage{xspace}
\usepackage{rotating,tabularx}
\usepackage{amsthm}
\usepackage{enumitem}
\usepackage[group-separator={,}]{siunitx}
\usepackage{multirow}
\usepackage{subcaption}
\usepackage{bbding}
\usepackage{makecell}
\usepackage{balance}
\usepackage{threeparttable}

\newcommand{\hide}[1]{}

\newcommand{\benchname}{OAG-Bench\xspace}
\newcommand{\contestname}{OAG-Challenge\xspace}
\newcommand{\datasetnum}{$20$\xspace}
\newcommand{\newdatasetnum}{ten\xspace}
\newcommand{\methodnum}{$70+$\xspace}
\newcommand{\expnum}{$120+$\xspace}
\newcommand{\vpara}[1]{\vspace{0.07in}\noindent\textbf{#1 }}
\newcommand{\todo}[1]{\textbf{\color{red}[(TODO: #1 )]}}  

\newcommand{\modeloaglstm}[0]{LinKG$_L$\xspace}
\newcommand{\modeloagcnn}[0]{LinKG$_C$\xspace}
\newcommand{\modeloaggat}[0]{LinKG$_G$\xspace}

\newcommand{\beq}[1]{\vspace{-0.1in}\begin{equation}#1\end{equation}\vspace{-0.1in}}

\theoremstyle{problem}
\newtheorem{problem}{Problem}[section]

\copyrightyear{2024}
\acmYear{2024}
\setcopyright{rightsretained}
\acmConference[KDD '24]{Proceedings of the 30th ACM SIGKDD Conference on Knowledge Discovery and Data Mining}{August 25--29, 2024}{Barcelona, Spain}
\acmBooktitle{Proceedings of the 30th ACM SIGKDD Conference on Knowledge Discovery and Data Mining (KDD '24), August 25--29, 2024, Barcelona, Spain}\acmDOI{10.1145/3637528.3672354}
\acmISBN{979-8-4007-0490-1/24/08}

\begin{document}

\title{\benchname: A Human-Curated Benchmark for \\ Academic Graph Mining}



\author{Fanjin Zhang}
\authornote{Part of the work was done when Fanjin worked at Zhipu AI.}
\email{fanjinz@tsinghua.edu.cn}
\orcid{0000-0001-8551-1966}
\affiliation{\institution{Tsinghua University}}

\author{Shijie Shi}
\email{shishijie2020@163.com}
\orcid{0009-0000-0597-7337}
\affiliation{\institution{Zhipu AI}}

\author{Yifan Zhu}
\email{yifan_zhu@bupt.edu.cn}
\orcid{0000-0002-7695-1633}
\affiliation{\institution{Beijing University of Posts and Telecommunications}}

\author{Bo Chen}
\email{cb21@mails.tsinghua.edu.cn}
\orcid{0000-0002-9629-5493}
\affiliation{\institution{Tsinghua University}}

\author{Yukuo Cen}
\email{yukuo.cen@zhipuai.cn}
\orcid{0000-0001-5682-2810}
\affiliation{\institution{Zhipu AI}}

\author{Jifan Yu}
\email{yujifan@tsinghua.edu.cn}
\orcid{0000-0003-3430-4048}
\affiliation{\institution{Tsinghua University}}

\author{Yelin Chen}
\email{ylin@stu.xju.edu.cn}
\orcid{0000-0002-9461-2889}
\author{Lulu Wang}
\email{wanglulu@stu.xju.edu.cn}
\orcid{0000-0001-5584-2908}
\author{Qingfei Zhao}
\email{zhaoqingfei21@mails.ucas.ac.cn}
\orcid{0009-0007-5962-2997}
\affiliation{\institution{Zhipu AI}}

\author{Yuqing Cheng}
\email{chengyuqing@mail.ccom.edu.cn}
\orcid{0009-0003-3634-1202}
\author{Tianyi Han}
\email{tianyi.han@aminer.cn}
\orcid{0009-0001-5829-6030}
\affiliation{\institution{Zhipu AI}}

\author{Yuwei An}
\email{ayw.sirius19@gmail.com}
\orcid{0009-0006-6866-9589}
\author{Dan Zhang}
\email{zd21@mails.tsinghua.edu.cn}
\orcid{0000-0003-1115-3945}
\affiliation{\institution{Tsinghua University}}

\author{Weng Lam Tam}
\email{rainatam9784@gmail.com}
\orcid{0009-0008-4517-3593}
\author{Kun Cao}
\email{bzsy2476203449@gmail.com}
\orcid{0009-0001-2004-4006}
\author{Yunhe Pang}
\email{pangyunhe@ncepu.edu.cn}
\orcid{0009-0005-2425-8836}
\affiliation{\institution{Zhipu AI}}

\author{Xinyu Guan}
\email{guanxinyu@gmail.com}
\orcid{0009-0000-6901-7925}
\affiliation{\institution{Biendata}}

\author{Huihui Yuan}
\email{huihui.yuan@aminer.cn}
\orcid{0009-0001-1936-6323}
\author{Jian Song}
\email{jian.song@aminer.cn}
\orcid{0000-0001-9024-4954}
\author{Xiaoyan Li}
\email{xinyan.li@aminer.cn}
\orcid{0009-0002-6124-4204}
\affiliation{\institution{Zhipu AI}}

\author{Yuxiao Dong}
\email{yuxiaod@tsinghua.edu.cn}
\orcid{0000-0002-6092-2002}
\affiliation{\institution{Tsinghua University}}

\author{Jie Tang}
\authornote{Jie Tang is the corresponding author.}
\email{jietang@tsinghua.edu.cn}
\orcid{0000-0003-3487-4593}
\affiliation{\institution{Tsinghua University}}

\hide{
\author{Ben Trovato}
\authornote{Both authors contributed equally to this research.}
\email{trovato@corporation.com}
\orcid{1234-5678-9012}
\author{G.K.M. Tobin}
\authornotemark[1]
\email{webmaster@marysville-ohio.com}
\affiliation{%
  \institution{Institute for Clarity in Documentation}
  \streetaddress{P.O. Box 1212}
  \city{Dublin}
  \state{Ohio}
  \country{USA}
  \postcode{43017-6221}
}
}


\makeatletter
\def\@ACM@checkaffil{
    \if@ACM@instpresent\else
    \ClassWarningNoLine{\@classname}{No institution present for an affiliation}%
    \fi
    \if@ACM@citypresent\else
    \ClassWarningNoLine{\@classname}{No city present for an affiliation}%
    \fi
    \if@ACM@countrypresent\else
        \ClassWarningNoLine{\@classname}{No country present for an affiliation}%
    \fi
}
\makeatother

\hide{
\author[Zhang et al.]{
    Fanjin Zhang$^{1,2 *}$\authornote{Part of the work was done when Fanjin worked at Zhipu AI.}, 
    Shijie Shi$^{2}$, Yifan Zhu$^{1}$, 
    Bo Chen$^{1}$, Yukuo Cen$^{2}$, Jifan Yu$^{1}$, Yelin Chen$^{2}$, 
    Lulu Wang$^{2}$, \\ Qingfei Zhao$^{2}$, 
    Yuqing Cheng$^{2}$, Tianyi Han$^{2}$,
    Yuwei An$^{1}$, Dan Zhang$^{1}$, Weng Lam Tam$^{2}$, 
    Kun Cao$^{2}$, Yunhe Pang$^{2}$, 
    Xinyu Guan$^{3}$, Huihui Yuan$^{2}$, Jian Song$^{2}$, Xiaoyan Li$^{2}$, 
    Yuxiao Dong$^{1}$, 
    Jie Tang$^{1}$
}
\affiliation{%
  \institution{$^1$ Department of Computer Science and Technology, Tsinghua University, Beijing, China}
}
\affiliation{
    $^2$ Zhipu AI, Beijing, China $^3$ Biendata, Beijing, China
}

\email{
  open-academic-graph@googlegroups.com
}

}



\renewcommand{\authors}{Fanjin Zhang, Shijie Shi, Yifan Zhu, 
    Bo Chen, Yukuo Cen, Jifan Yu, Yelin Chen, 
    Lulu Wang, Qingfei Zhao, 
    Yuqing Cheng, Tianyi Han,
    Yuwei An, Dan Zhang, Weng Lam Tam, 
    Kun Cao, Yunhe Pang, 
    Xinyu Guan, Huihui Yuan, Jian Song, Xiaoyan Li, 
    Yuxiao Dong, Jie Tang}

\renewcommand{\shortauthors}{Fanjin Zhang et al.}

\begin{abstract}
  With the rapid proliferation of scientific literature,
  versatile academic knowledge services increasingly rely on comprehensive academic graph mining.
  Despite the availability of public academic graphs, benchmarks, and datasets, 
  these resources often fall short in multi-aspect and fine-grained annotations,
  are constrained to specific task types and domains,
  or lack underlying real academic graphs.
  In this paper, we present \benchname,
  a comprehensive, multi-aspect, and fine-grained human-curated benchmark based on 
  the Open Academic Graph (OAG).
  \benchname covers 10 tasks, \datasetnum datasets, \methodnum baselines, and \expnum experimental results to date.
  We propose new data annotation strategies for certain tasks
  and offer a suite of data pre-processing codes, algorithm implementations, and standardized evaluation protocols to facilitate academic graph mining.
  Extensive experiments reveal that even advanced algorithms like large language models (LLMs) 
  encounter difficulties in addressing key challenges in certain tasks, 
  such as paper source tracing and scholar profiling.
  We also introduce 
  the Open Academic Graph Challenge (\contestname) 
  to encourage community input and sharing. 
  We envisage that \benchname can serve as a common ground for the community to evaluate and compare algorithms in academic graph mining,
  thereby accelerating algorithm development and advancement in this field.
  \benchname is accessible at \url{https://www.aminer.cn/data/}.

\end{abstract}


\hide{


}




\maketitle

\input{intro.tex}

\input{background.tex}
\input{bench-overview.tex}

\input{task-detail.tex}

\input{leaderboard.tex}
\input{conclusion.tex}



\begin{acks}
This work is supported by Natural Science Foundation of China (NSFC) 62425601 and  62276148, Technology and Innovation Major Project of the Ministry of Science and Technology of China under Grant 2020AAA0108400, the New Cornerstone Science Foundation through the XPLORER PRIZE. 
We also thank Weibin Liao, Chao Yu, Kai Yu, and Zheng Jiang for their contribution to code reproducibility.
\end{acks}


\newpage

\bibliographystyle{ACM-Reference-Format}
\bibliography{ref}

\appendix


\input{appendix.tex}

\end{document}

%% file: intro.tex
\section{Introduction}

\hide{
The overarching goal of building accurate academic knowledge graphs is to 
deepen our comprehension of the development, nature, and trends of science.
Platforms such as
Google Scholar\footnote{\url{https://scholar.google.com/}}, 
Web of Science\footnote{\url{https://www.webofscience.com/}}, 
and Semantic Scholar\footnote{\url{https://www.semanticscholar.org/}}
have established their own academic graphs,
offering specialized knowledge services.
For instance, AMiner\footnote{\url{https://www.aminer.cn/}}~\cite{tang2008arnetminer}, 
an academic search and mining system, offers services such as paper recommendation, trend analysis, and academic ranking. 
There have been more than $16$ million 
annual page visits
on AMiner,
underscoring the significant demand
of researchers for academic knowledge services.
}

The overarching goal of academic data mining is to 
deepen our comprehension of the development, nature, and trends of science.
It offers the potential to unlock enormous scientific, technological, and educational value~\cite{wang2021science}.
For example, deep mining from academic data can assist governments in making scientific policies,
support companies in talent discovery,
and help researchers acquire new knowledge more efficiently.

\hide{
The popularization of digital publishing and big data technology advancements have made it feasible to semi-automatically construct large-scale academic knowledge graphs using machine learning techniques. 
Despite the surge in online publications,
there remains a notable absence of a standardized benchmark to evaluate various algorithms for
constructing academic knowledge graphs
and serving academic knowledge applications.
}

\begin{table}
	\centering
	\caption{Comparison between academic knowledge graphs (AKG) and academic benchmarks. \textmd{Biomed: biomedicine.}} \label{tb:bench_compare}
\begin{tabular}{@{}lccccc@{}}
	\toprule[1.2pt]
	\makecell[c]{AKG /\\Benchmark}  & \makecell[c]{Multiple\\Tasks}   & Domain     & \makecell[c]{Task\\Type}  & \makecell[c]{Baseline\\Codes} &   \makecell[c]{Leader-\\board}         \\ \midrule
	MAG~\cite{sinha2015overview} & -     & All & -          & -          & -             \\
	OAG~\cite{zhang2019oag,zhang2022oag} & - & All & - & - & - \\
	AceKG~\cite{wang2018acekg} & Partial & All & Graph & - & - \\
	S2ORC~\cite{lo2020s2orc} & \Checkmark & All & NLP & Partial & \Checkmark \\
	BLURB~\cite{gu2021domain} & \Checkmark & Biomed. & NLP & \Checkmark & \Checkmark \\
	\midrule
	\makecell[c]{\benchname}  & \Checkmark & All & Diverse & \Checkmark & \Checkmark  \\ \bottomrule[1.2pt]
\end{tabular}
\end{table}

The landscape of academic data mining is rich with entity-centric applications, such as
paper recommendation, expert finding, and venue recommendation.
Several popular academic mining systems,
such as Semantic Scholar\footnote{\url{https://www.semanticscholar.org/}},
ResearchGate\footnote{\url{https://www.researchgate.net/}},
and AMiner\footnote{\url{https://www.aminer.cn/}},
are all powered by academic knowledge graphs (\textbf{AKG})\footnote{In this paper, we use \textit{academic knowledge graph} and \textit{academic graph} interchangeably.}.
Based on different data sources,
there have been multiple public academic graphs and academic benchmarks,
such as Microsoft Academic Graph (MAG)~\cite{sinha2015overview} and S2ORC~\cite{lo2020s2orc}.
A comparative overview of these academic resources is presented in
Table \ref{tb:bench_compare}.
However, there remain several defects in existing popular datasets
that may hinder promising explorations,
which are summarized as follows:

\begin{itemize}[leftmargin=*]
	\item Public academic graphs, such as MAG and OAG~\cite{zhang2019oag},
	lack multi-aspect and fine-grained annotations,
	impeding potential evaluation of downstream tasks on top of them.
	\item Academic benchmarks, such as S2ORC and BLURB~\cite{gu2021domain},
	are limited to specific task types (e.g., NLP tasks) and domains (e.g., biomedicine),
	which may not cover the full spectrum of academic tasks, such as various graph-based tasks.
	\item Separate academic datasets, such as PubMedQA~\cite{jin2019pubmedqa} and concept taxonomy datasets~\cite{shen2020taxoexpan},
	often do not include or align with large-scale and comprehensive academic graphs,
 resulting in a divergence from real-world scenarios.
\end{itemize}

\begin{figure}[t]
	\centering
\begin{minipage}{8cm}
\renewcommand{\thempfootnote}{\arabic{mpfootnote}}
\centering
	\includegraphics[width=7.5cm]{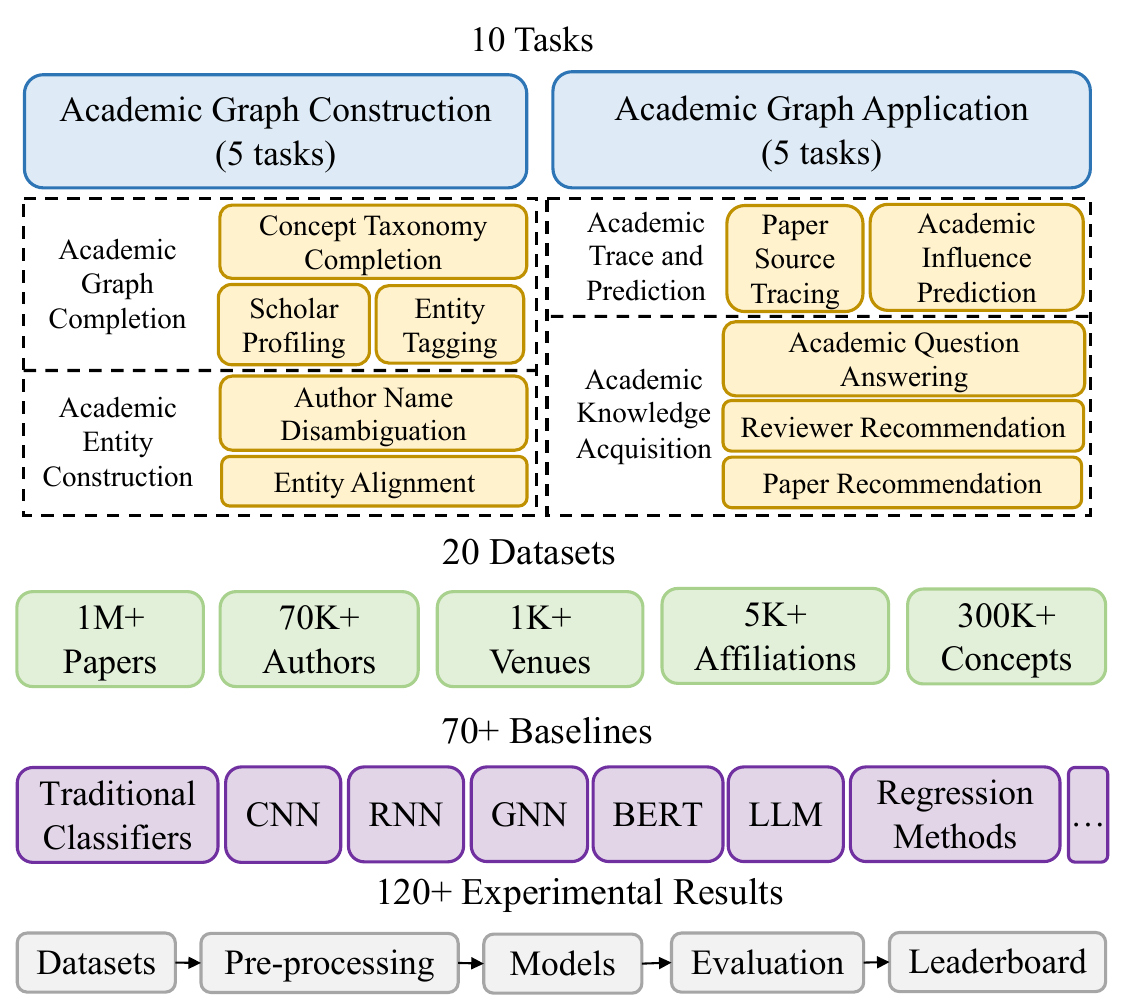}	
	\caption[Compact Routing Example]%
    {\benchname overview. 
 }
	\label{fig:oag-bench-overview}
\end{minipage}
\end{figure}

\vpara{Present Work.}
To this end, we introduce \benchname, 
a meticulously human-annotated academic benchmark for academic graph mining.
\benchname currently includes ten tasks, \datasetnum datasets,
\methodnum baseline methods, and \expnum experimental results.
Figure \ref{fig:oag-bench-overview} provides an overview of \benchname.
Specifically, 
\begin{enumerate}[leftmargin=*]
\hide{
\item The academic tasks are divided into two types: academic graph construction
and academic applications.
Academic graph construction constructs, completes, or corrects \todo{basic} 
relationships between entities,
such as paper citation relationships and auther-auther disambi....
Academic applications study the cognitive impact of authors or researches.
Worth mention that academic applications can be empowered by AKG because accurate....
}
\item For the \textbf{design principles} of \benchname,
we aim to conduct comprehensive and fine-grained annotations on 
the large-scale OAG for the full life cycle of academic graph mining.
Firstly, we annotate the nodes and edges of the academic knowledge graph
and identify valuable and challenging tasks during this process,
such as author name disambiguation.
Then, 
powered by the academic graph,
academic applications explore tasks beyond the academic graph itself
and study knowledge acquisition and cognitive impact, such as paper source tracing (C.f. Section \ref{sec: framework}).
\item For the \textbf{datasets} in \benchname,
we construct various human-curated datasets for diverse tasks.
We also propose new annotation strategies for certain tasks,
such as checking inconsistent paper assignments across sources for incorrect paper-author assignment detection
and marking the sources of papers via online paper reading groups.
Notably, \newdatasetnum datasets in eight tasks are newly constructed.
The dataset sizes in \benchname range from thousands to millions.
\item For the \textbf{evaluation} of \benchname,
\benchname provides corresponding data processing methods,
evaluation metrics, and at least three baseline methods for each task.
\benchname implements a wide range of methods,
covering traditional machine learning methods, shallow convolutional/recurrent/graph neural networks, 
LLMs, etc.
Experimental investigations show that advanced generation-based LLMs hold promise in some tasks like author name disambiguation,
but they still struggle with tasks like scholar profiling and paper source tracing.
\end{enumerate}

To sum up, \benchname makes the following contributions: First, 
 we provide multi-aspect and fine-grained human-curated datasets 
 that cover the full life cycle of academic graph mining.
	Second, we release a series of data pre-processing codes, algorithm implementations, 
	and standardized evaluation protocols to assist researchers
 in getting started quickly in academic graph mining.
	Finally,  based on \benchname, interested researchers or practitioners can 
	develop advanced AKG-based algorithms, 
	study the foundation models for academic graph mining, and so forth.

%% file: background.tex
\section{Background}

This section first gives the formal definition of academic knowledge graphs
and then introduces related academic datasets.

\subsection{Academic Knowledge Graph}

An academic knowledge graph (AKG) is defined as a graph $AKG=\{E, R\}$ where each entity $e\in E$ and each relation $r\in R$ are associated with type mapping functions $\tau(e): E \to C$ and $\phi(r): R \to D$, respectively. 
$C$ and $D$ represent the sets of entity and relation types with $|C| > 1$ and $|D| > 1$. 
Each entity pair $e_1$ and $e_2$ is linked by a specific relation $r \in R$ to form a tuple $(e_1, r, e_2)$.

For instance, an academic graph is a heterogeneous entity graph 
that encompasses multiple types of entities, 
such as authors, papers, and venues. 
The relation set, represented by $D$, 
includes several key relationships:
the \textit{authorship} relation, which connects authors and papers;
the \textit{paper-publish-in-venue} relation that links papers to the venues where they are published;
the \textit{co-authorship} relation, indicating collaborations between authors, etc.

\subsection{Academic Datasets}
Some organizations have made their academic graphs available,
including MAG~\cite{sinha2015overview}, OAG~\cite{zhang2019oag,zhang2022oag},
AceKG~\cite{wang2018acekg}, OpenAlex\footnote{\url{https://openalex.org/}},
and CrossRef\footnote{\url{https://www.crossref.org/}}.
These graphs are typically large-scale,
but are rarely carefully annotated to benchmark a wide range of academic tasks.
Additionally, some benchmarks based on academic corpus have been proposed,
such as S2ORC~\cite{lo2020s2orc}, SciDocs~\cite{cohan2020specter}, and BLURB~\cite{gu2021domain},
but they mainly target NLP tasks and 
overlook the intricate structure of academic graphs.

To bridge the gap,
our objective is to meticulously annotate large-scale academic graphs
to benchmark various tasks for academic graph mining.
Our initiative, \benchname, leverages the
Open Academic Graph (OAG)\footnote{\url{https://www.aminer.cn/open-academic-graph}},
which was initially generated by linking two large academic graphs: MAG and AMiner.
OAG aligned large-scale entities in MAG and AMiner,
including papers, authors, affiliations, and venues,  with an accuracy of over $97\%$.
 It has made available the alignment relations between these two graphs alongside their metadata. 
As MAG turned down its service at the end of 2021,
OAG has expanded its data sources to include
PubMed, ArXiv, CrossRef, and so forth.
To date, five versions of OAG have been released,
amassing around 700 million entities and 2 billion relations.

%% file: bench-overview.tex
\section{\benchname Framework}
\label{sec: framework}

In this section, we first propose the overall design principle of \benchname,
and then present the detailed workflow about how to construct comprehensive and high-quality datasets.

\begin{figure}[t]
	\centering
	\includegraphics[width=8cm]{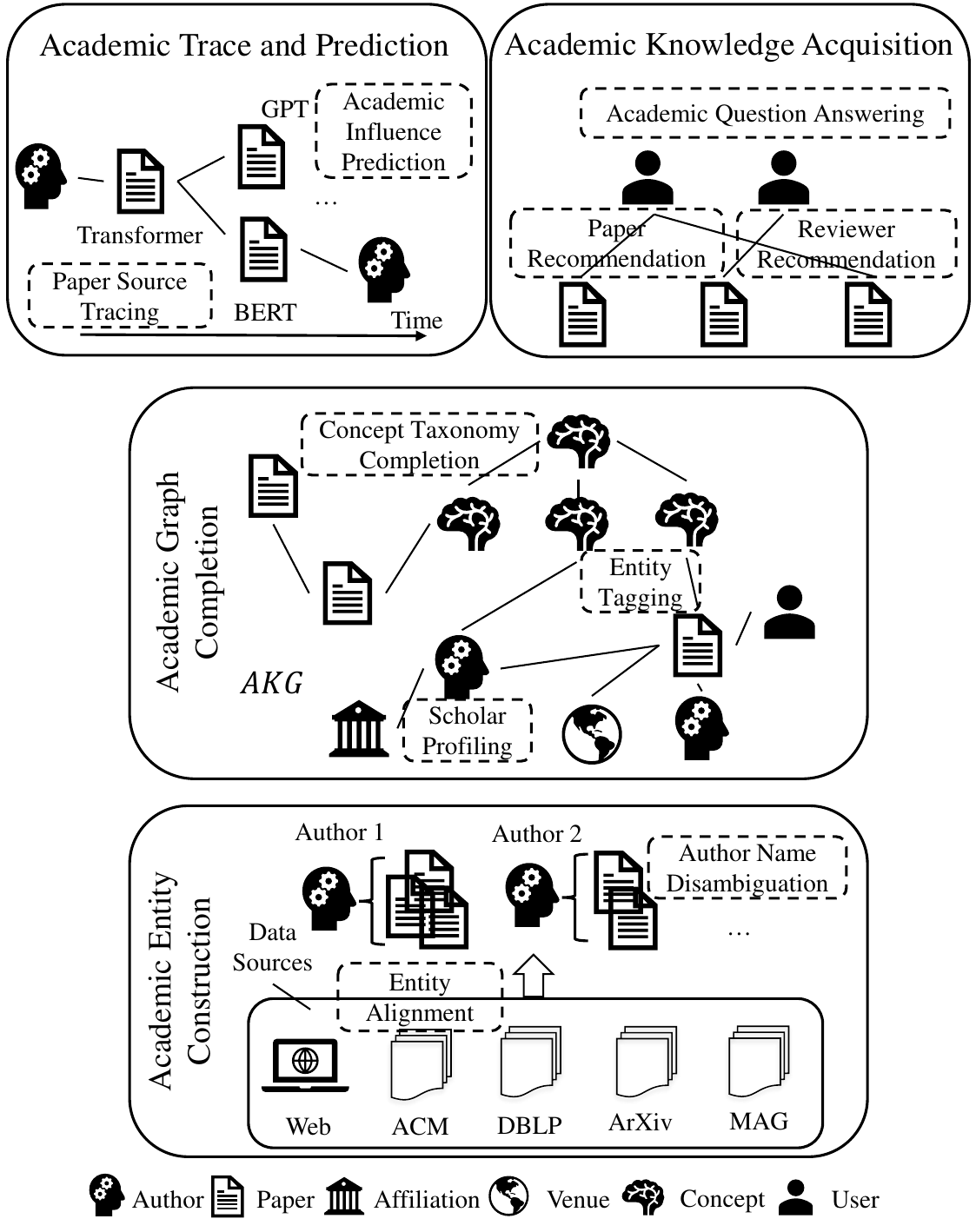}	
	\caption{
        The overall construction framework of \benchname.
    }
	\label{fig:oag-bench-framework}
\end{figure}

As depicted in Figure \ref{fig:oag-bench-framework},
we host a series of data collection and annotation efforts 
to conduct multi-aspect and fined-grained labeling based on the OAG.
The framework aims to leverage high-quality academic knowledge graphs (AKG)
to facilitate academic data mining.
Therefore, the framework is structured into two types of tasks:
AKG construction and AKG-empowered academic applications.
AKG construction focuses on disambiguating or enriching graph nodes and
correcting or completing graph edges,
consisting of \textit{Academic Entity Construction} and \textit{Academic Graph Completion}.
Beyond basic academic relationships, Academic applications delve into knowledge and cognition,
consisting of \textit{Academic Knowledge Acquisition} and \textit{Academic Trace and Prediction}.

(1) \textit{Academic Entity Construction.}
The construction of academic entities is fundamental to the construction of academic graphs.
This stage mainly identifies the identical real-world entities across data sources.
For notoriously ambiguous entities, i.e., authors,
we further incorporate the author name disambiguation task.

(2) \textit{Academic Graph Completion.}
Building upon the conflated academic entities,
this stage aims to establish connections between different entities to complete and enrich academic graphs.
Specifically, we engage in fine-grained scholar profiling labeling
and attach concepts to entities, such as authors, papers, and concepts.

(3) \textit{Academic Knowledge Acquisition.}
On top of high-quality academic graphs,
this stage focuses on the acquisition of academic knowledge and 
models the multifaceted relations between users and papers.
We gather user behavior records from real academic systems to build corresponding datasets.

(4) \textit{Academic Trace and Prediction.}
Besides the correlation between academic knowledge and users,
this stage aims to further explore the cognitive influence exerted by papers and authors. 
It involves retrospective analysis to pinpoint the pivotal references that have inspired a research paper.
Looking forward, the challenge lies in forecasting impactful papers or authors.

Table \ref{tb:bench_overview} summarizes the specifics of the datasets in \benchname.
\benchname includes diverse tasks and datasets
since 
the construction of academic graphs is complex and can not be conducted end-to-end.
Furthermore, the applications of academic graphs also involve 
diverse paper-centric, author-centric, and user-centric services.
Although \benchname currently includes ten different tasks,
these tasks could facilitate each other.
For instance, profiling scholars precisely can 
help to attach concept tags to scholars.
These tasks can also foster other academic tasks.
For example, paper recommendation datasets are also
valuable assets for similar paper search.
In the following, we will present 
the design choices of the tasks in each module,
corresponding task definitions,
and the construction methods of related datasets.

\begin{table*}
    \newcolumntype{C}{>{\centering\arraybackslash}p{3cm}}
    \small
    \centering
     \begin{threeparttable}
    \caption{Dataset overview in \benchname.
    \textmd{The format of \textit{\#Datasets} column is \textit{\#Datasets/\#New datasets} for each task.}
    } \label{tb:bench_overview}
    \begin{tabular}{p{3.5cm}<{\centering} p{2.5cm}<{\centering} p{1cm}<{\centering} p{2cm}<{\centering} p{1cm}<{\centering} p{5cm}<{\centering}}
        \toprule
        Task & Data Source & 
        \#Datasets
        & Data Volume & \#Baselines & 
        Data characteristics  \\
        \midrule
        Entity alignment & AMiner/DBLP/MAG & 
        3/2 & 1K-10K & 5 &  Matching heterogeneous entities\\
        Author name disambiguation & AMiner &
        3/1 & 1M & 10 & Million-scale human-annotated data\\
        Scholar profiling  & AMiner & 
        2/1 & 2K-9K &  13 & Long attribute extraction for long texts\\
        Entity tagging  & AMiner & 
        2/1 & 11K-900K & 12 & A large number of class labels \\
        Concept taxonomy completion & MAG/AMiner &
        3/1 & 1K-300K & 3 & Professionally-labeled data for the AI field\\
        Paper recommendation & AMiner &
        1/0 & 10K & 4 & User click records in a real academic system\\
        Reviewer recommendation  & Frontiers &
        1/1 & 200K & 4 & Authentic public review records \\
        Academic question answering & Zhihu/StackExchange & 
        1/0 & 18K & 6 & Automatic QA for academic domain\\
        Paper source tracing & AMiner &
        1/1 & 2K & 11 & Careful annotations by researchers\\
        Academic influence prediction & AMiner & 
        3/2 & 1K-1M & 12 & Summarizing Test-of-Time award in CS\tnote{\textasteriskcentered}\\
        \bottomrule

    \end{tabular}
\footnotesize
\begin{tablenotes}
\item[\textasteriskcentered] CS: computer science.
\end{tablenotes}

\end{threeparttable}

\end{table*}

\subsection{Academic Entity Construction}

To integrate academic data from multiple sources,
various types of entities need to be aligned.
Thus, we first include the \textit{entity alignment} task.
In view of the severe ambiguity of author names,
we further add the \textit{author name disambiguation} task.

\vpara{Entity Alignment.}
Given two entity sets $E_1$ and $E_2$,
the goal of entity alignment is to generate entity matchings 
$L = \{(e_1, e_2) | e_1 \in E_1, e_2 \in E_2\}$ such that 
$e_1$ and $e_2$ refer to the same real-world entity.
Specifically, we consider three types of entities, i.e., authors, affiliations, and venues.
As for dataset annotation,
we randomly sample a venue set and an affiliation set,
and then manually label venue pairs with high similarity calculated by the Jaccard Index,
and construct affiliation alignment pairs by using aliases or former names derived from the information box of their Wikipedia entries.
We utilize Wikipedia due to its high data quality.
Meanwhile, it allows us to accurately obtain positive affiliation alignment pairs without the need for manual labeling.
For author alignment, we sample top-viewed computer science authors from AMiner,
and then manually pair them with DBLP authors according to 
their affiliations, published venues, and papers.
As a result, we construct $1{\small,}200$ venue pairs,
$5{\small,}000$ affiliation pairs, and \num{10000} author pairs.

\vpara{Author Name Disambiguation (AND).} Aiming to disambiguate the same-name authors,
AND is a key and challenging task in academic knowledge graph construction.
We adopt the WhoIsWho~\cite{chen2023web} dataset,
a million-scale human-annotated dataset for author name disambiguation.
WhoIsWho breaks down the task into three subtasks:
(1) From-scratch Name Disambiguation (SND),
(2) Real-time Name Disambiguation (RND), and
(3) Incorrect Assignment Detection (IND).
While existing research primarily concentrates on SND and RND, the IND task has received less attention despite its growing importance with the expansion of academic databases.
Given an author profile with paper lists,
IND aims to detect incorrectly assigned papers to this author.
To address the IND challenge, 
if we were to randomly select author profiles for annotation of their paper assignments, there's a high likelihood that we would encounter numerous profiles with a little ambiguity.
Thus, we propose \textit{an effective cross-checking annotation strategy}.
Specifically, we utilize existing paper alignments and author alignments between AMiner and DBLP, 
and then gather inconsistent paper-author assignments for further expert checking. 
This strategy ensures that the profiles under checking have a high likelihood of inaccuracy (with a significant error rate exceeding 30\% for these inconsistencies within AMiner). 
Subsequently, all papers associated with AMiner authors that have incorrect assignments are manually checked by experts.
Finally, the refined IND dataset includes $1{\small,}691$ authors, \num{326738} papers,
with an assignment error rate of $11.32\%$
and reaching $1.5$ times the scale of papers of the IND task in WhoIsWho.

\subsection{Academic Graph Completion}
Academic graph completion aims to enrich academic graphs from two aspects --- entities and relations.
To enrich entities, we include the \textit{scholar profiling} task to extract multidimensional attributes for authors.
To enrich relations, we include the \textit{entity tagging} task to attach concepts to entities.
Concepts are abstract entities that can endow semantics to entities.
To further build a hierarchical knowledge structure,
we also include the \textit{concept taxonomy completion} task to identify hypernyms and hyponyms for new concepts.

\vpara{Scholar Profiling.}
Profiling scholars from big data is a vital task in scholar mining,
and it becomes harder and harder due to 
data fragmentation, modeling lengthy texts, data noise, etc.
Previous works on scholar profiling usually extract attributes from scholars' homepages or search engines.
In \benchname, besides profiling scholars from search engines, 
we introduce a new complex setting ---
\textit{Multidimensional Scholar Profiling from Long Texts},
which aims to extract multiple attributes in lengthy texts.
Each attribute extraction includes the starting and ending positions in the text. 
Importantly, long attributes are also taken into consideration, such as work experience and education experience.
These attributes can often exceed 100 tokens. 
Traditional scholar profiling or named entity recognition tasks seldom focus on extracting such long attributes.
For data annotation, scholars with detailed biographical descriptions are randomly sampled.
Then, we manually label the starting and ending positions of each attribute in the texts.
Finally, we construct $2{\small,}099$ scholars with $12$ attributes.

\vpara{Entity Tagging.}
Aiming at associating entities with concept labels,
entity tagging is an important step in building semantic and hierarchical academic graphs. 
We introduce scholar interest extraction and paper topic classification 
to attach concepts to scholars and papers, respectively.
Scholar interest extraction aims to extract scholars' research interests from their publications.
Derived from 2017 Open Academic Data Challenge\footnote{\url{https://www.biendata.xyz/competition/scholar/}},
the dataset of this task contains manually annotated $789$ research interest tags for \num{11357} scholars 
and their papers. 
Paper topic classification aims to classify papers into several topics based on the paper citation network.
For dataset construction, based on the DBLP paper citation network\footnote{\url{https://originalstatic.aminer.cn/misc/dblp.v12.7z}},
each paper is assigned one of nine topics\footnote{\url{https://numbda.cs.tsinghua.edu.cn/~yuwj/TH-CPL.pdf}. The topics include 
high-performance computing, computer networks, network and information security,
theoretical computer science, system software and software engineering, database and data mining,
artificial intelligence and pattern recognition, computer graphics and multimedia, human-computer interaction, and pervasive computing.
}
related to computer science based on its publication venue. 

\vpara{Concept Taxonomy Completion.}
Concept taxonomies are typically manually created by experts, 
like defining ``deep learning'' falls under ``machine learning''. 
The automatic construction of concept taxonomies is a critical challenge 
in the fast-evolving landscape of knowledge concepts, 
which is beneficial to the organization of knowledge in the realm of big data.
Given an existing concept hierarchy tree (Taxonomy) $T_0$ 
and a set of new concepts $C$, the goal of concept taxonomy completion is to predict its hypernym $pa(c) \in T_0$ and hyponym $ch(c) \in T_0$ for each new concept $c \in C$ to complete and expand the existing concept hierarchy tree.
We adopt two MAG taxonomies~\cite{sinha2015overview} (MAG-Full and MAG-CS) as two taxonomy datasets.
These two datasets are large-scale but not carefully verified by experts.
Thus, we introduce a newly manually curated dataset covering AI sub-fields by AI researchers, 
with $1{\small,}335$ concepts and $1{\small,}283$ edges. 
The guidelines of edge construction refer to relevant textbooks and the ACM Computing Classification System.

\subsection{Academic Knowledge Acquisition}

Academic services based on academic graphs provide convenience for researchers to acquire knowledge actively or passively. 
For passive academic recommendation, we include \textit{paper recommendation} and \textit{reviewer recommendation}.
For active knowledge acquisition, we include \textit{academic question answering} task.

\vpara{Paper Recommendation.}
As the volume of papers surges, 
researchers face increasing challenges in locating relevant literature. 
Given a user-paper bipartite graph $G = \{U, P, R\}$, 
where $U$ is the user set, $P$ is the paper set, 
and $R$ signifies interactions (e.g., clicks) between users and papers, 
the goal of paper recommendation is to predict the next paper a user will interact with~\cite{zhang2023apegnn}.
We collect user behavior data based on the real AMiner system.
AMiner provides a real-time paper recommendation service for researchers on the homepage.
Researchers can offer several keywords to subscribe to relevant research papers.
The back-end recommendation engine makes recommendations based on the users' historical click records.
This dataset includes $5{\small,}340$ users, \num{14967} papers, and \num{163084} interactions as of October 2021. 
To ensure quality, only users/papers with over $10$ clicks/be-clicked instances are included.

\vpara{Reviewer Recommendation.}
As the volume of submissions to academic journals and conferences increases,
reviewer recommendation becomes increasingly hard.
Different from paper recommendations, 
reviewer recommendation aims to pair papers with proficient and willing reviewers.
Given a paper submission set $S$, a reviewer set $A$, and 
known paper-reviewer matches $R \subseteq S \times A$, 
this task is to predict the reviewer $a \in A$
for a new submission record $s_i \in S$, 
Additional information, including paper metadata and reviewer expertise, is available.
For data collection, we extract real paper-reviewing records from the open-access platform Frontiers.
After processing, it includes \num{210069} reviewers and \num{225478} papers,
with each paper having at least $2$ reviewers. 
Furthermore, we match reviewers to authors in OAG using names, affiliations, and research interests, linking approximately half of the reviewers to the OAG. These reviewers are associated with their respective publications.

\vpara{Academic Question Answering.}
Traditional keyword-based information retrieval cannot satisfy professional knowledge retrieval in the era of artificial intelligence. 
For instance, consider the question, 
``Can neural networks be used to prove conjectures?''.
How to retrieve answers and evidence from scholarly literature? 
Given an academic question $q$ and a paper set $P^q = \{p^q_1, p ^q_2,..,p^q_N\}$,
the goal of academic question answering is to select the most relevant papers
from the candidate set $P^q$.
We adopt \textbf{OAG-QA}~\cite{tam2022parameter} dataset,
which is derived from academic question-answering platforms. 
We retrieve question posts from StackExchange and Zhihu websites, 
extract the paper URL mentioned 
in the answer, 
and match it with the paper in OAG~\cite{zhang2019oag}. 
It comprises \num{17948} question-paper pairs. 
Questions cover $22$ disciplines and $87$ topics, 
forming a two-level hierarchical structure; that is, each topic belongs to a discipline. 
For each topic, \num{10000} candidate papers, 
including the ground-truth papers in the answers, are included.

\subsection{Academic Trace and Prediction}
Understanding the evolution of science on the cognitive level offers the potential to predict, change, and finally invent the future.
Tracing back to the past, we include \textit{paper source tracing} task to identify the sources of research papers.
To predict future potential academic impact, 
we include two tasks, i.e.,  \textit{paper influence prediction} and \textit{author influence prediction}.

\vpara{Paper Source Tracing (PST).}
Tracing the sources of papers is crucial for understanding technological essence and uncovering innovation patterns.
Given a paper $p$ (including its full text) and its references, 
the goal of PST is to identify the most important references (termed \textit{ref-source}) 
that largely inspired the paper $p$ in terms of ideas or methods. 
The source papers of a given paper are defined by the following principles:
(1) the main idea of the paper $p$ is inspired by the reference; 
or (2) the main method of the paper $p$ comes from the reference. 
In other words, this paper would not come into being without these source papers. 
We carefully build a dataset \textbf{PST-Bench} for this task.
Given the specialized knowledge required for paper source tracing, 
dozens of computer science graduate students were employed to mark the sources of papers in their familiar fields. 
The annotation process was organized in an online paper reading group,
where each student needed to share two papers and mark their source papers each week.
After collection, expert-checking, and preprocessing, $2{\small,}141$ labeled computer science papers were obtained. 
We conducted a human evaluation on the test set, with senior researchers double-checking 100 papers. The accuracy rate was 94\%.

\vpara{Academic Influence Prediction.}
Paper influence prediction aims to forecast a paper's impact $\Delta yr$ years later 
based on its metadata and citation relationships.
Regarding the ``Test-of-Time Paper Award'' (TOT award) as an indicator of high impact,
we collate TOT awards 
in computer science venues\footnote{\url{https://numbda.cs.tsinghua.edu.cn/~yuwj/TH-CPL.pdf}}.
Awards with similar meanings include the Most Influential Award, Sustained Influence Award, etc. 
At present, a total of $1{\small,}063$ papers awarded by 2022 have been collected. 
Similarly, author influence prediction seeks to predict an author's influence 
$\Delta yr$ years later 
using their papers and citation relationships. 
For this task, we provide two datasets.
First, we adopt \textbf{AuthPred-2017}\footnote{\url{https://www.biendata.xyz/competition/Tsinghua_course3/}}.
This dataset contains a subset of AMiner authors and papers published by these authors before 2011, 
intending to predict
the citations of these authors as of 2016. 
This dataset uses AMiner's citation statistics
and provides \num{1112931} authors for training and \num{123823} authors for testing.
In addition, we construct a new dataset \textbf{AuthPred-2022}.
This dataset contains a subset of AMiner's authors in the field of computer science 
and papers published by these authors before 2017, 
with the goal of predicting the citations of these authors as of early 2022 (as calculated by Google Scholar). 
This dataset contains \num{26797} AMiner authors with Google Scholar links.

%% file: task-detail.tex
\section{Task Evaluations}

This section delves into representative tasks of each module in \benchname,
highlighting selected experiments.
Additional experiments are detailed in Appendix \ref{app:other-tasks}.
All codes have been available\footnote{\url{https://github.com/zfjsail/OAG-Bench}}.

\input{chapter/name-disambiguation.tex}
\input{chapter/scholar-profiling.tex}
\input{chapter/entity-tagging.tex}

\input{chapter/academic-recommendation.tex}
\input{chapter/academic-qa.tex}
\input{chapter/paper-source-tracing.tex}
\input{chapter/academic-influence-prediction.tex}

%% file: chapter/name-disambiguation.tex
\subsection{Author Name Disambiguation}

Since SND and IND are two widely studied tasks,
we take incorrect assignment detection (IND) as an illustration for evaluation.

\vpara{Baselines.}
We adopt graph-based anomaly detection methods and LLM-based methods as baselines.
For each author, graph-based methods first construct a paper similarity graph 
based on attribute similarity (e.g., co-authorship, co-organization) 
and then detect anomalies in the graph.
(1) Logistic Regression (\textbf{LR}):
injects top eigenvectors of each graph as features to perform node classification.
(2) \textbf{GCN}~\cite{kipf2017semi}:
employs graph convolutional networks as the encoder,
and then uses fully-connected layers to classify normal/abnormal nodes.
(3) \textbf{GCCAD}~\cite{chen2022gccad}: 
leverages graph contrastive learning and contrasts abnormal nodes with normal ones
in terms of their distances to the global context.
(4) \textbf{ChatGLM}~\cite{du2022glm}:
finetunes ChatGLM-6B model by inputting each author's paper list 
and asking the model whether one given paper is an anomaly or not.

\vpara{Evaluation Metrics.}
Due to the imbalance between positive and negative instances, 
we adopt the widely-used metric \textbf{AUC}.
Furthermore, we choose mean average precision (\textbf{MAP}) as another metric,
which pays more attention to the rankings of incorrect instances.
We take a macro average of each metric for each author.

\begin{table}[t]
    \centering
    \caption{
        Performance of incorrect assignment detection ($\%$).
    }
    \label{tb:exp-ind}
    \begin{tabular}{ccc}
        \toprule
        Method & AUC & MAP \\
        \midrule
        LR & 58.46 & 69.56 \\
        GCN & 62.48 & 71.18 \\
        GCCAD & 70.15 & 74.17 \\
        ChatGLM & \textbf{77.92} & \textbf{79.54}\\
        \bottomrule
    \end{tabular}
\end{table}

\vpara{Experimental Results.}
Table~\ref{tb:exp-ind} shows the performance of incorrect assignment detection.
We observe that graph neural network-based methods (GCN and GCCAD)
outperform the traditional method (LR) based on eigenvalue decomposition.
GCCAD explicitly contrasts abnormal paper nodes with other nodes,
yielding better performance than GCN.
Surprisingly, ChatGLM outperforms graph-based anomaly detection methods,
indicating the potential of the attention mechanisms in LLMs to 
capture the complex correlations between the target paper and the overall author profile.
The best performance of IND is not that satisfactory
compared with that of SND and RND tasks~\cite{chen2023web},
suggesting that more attention should be paid to the IND task 
for author name disambiguation in the future.

%% file: chapter/scholar-profiling.tex
\subsection{Scholar Profiling}

In this subsection, for entity attribute enrichment,
we present the evaluation of 
multidimensional scholar profiling from long texts. 

\vpara{Baselines.}
We select the latest NER methods based on pre-trained models:
(1) \textbf{Han et al.}~\cite{yan2022embarrassingly}:
use a Biaffine decoder 
to generate features for each start and end position 
and then employ CNN to classify locations based on spatial position dependence.
(2) \textbf{GlobalPointer}~\cite{su2022global}:
uses a multiplicative attention mechanism
to incorporate relative positional encodings of start and end positions
and alleviates class imbalance via modified loss functions.
(3) \textbf{UIE}~\cite{lu2022unified}:
is a generative pre-trained model based extraction framework 
with structure extraction languages and template-specific prompts.

\vpara{Evaluation Metrics.}
Precision, Recall, and F1 are computed by comparing predicted and annotated text segments for each attribute.
These individual attribute results 
are then averaged to obtain the overall evaluation result.

\begin{table}[t]
    \centering
    \caption{
        Extraction performance of multidimensional scholar profiling from long texts ($\%$).
    }
    \label{tb:profiling-long}
    \begin{tabular}{cccc}
        \toprule
        Method & Precision & Recall & F1 \\
        \midrule
        UIE & 43.14 & 35.86 & 39.15\\
        Global Pointer & 51.87 & 33.09 & 40.39 \\
        Han et al. & 50.33 & 43.76 & \textbf{45.09} \\
        \bottomrule
    \end{tabular}
\end{table}

\vpara{Experimental Results.}
\hide{
Table \ref{tb:profiling-basic} and 
Table \ref{tb:profiling-long}
display extraction results for search engine-based
and long text-based scholar profiling, respectively.
In Table \ref{tb:profiling-basic}, 
pre-trained models outperform traditional methods,
showcasing the expressive capacity and 
effectiveness of pre-trained models without manual feature design. 
BI-LSTM-CRF and partial pre-trained models exhibit similar performance in position extraction,
showing the suitability of both sequence labeling and pre-trained models. 
Large generative models excel in homepage extraction, 
though accuracy remains modest.
}
Table \ref{tb:profiling-long}
display extraction results for long text-based scholar profiling.
In the context of
scholar profiling from long texts,
Table \ref{tb:profiling-long} reveals that 
span-based methods like Han et al. surpass generation-based methods like UIE.
We also conduct preliminary experiments by giving some demonstrations and calling GPT-4~\cite{achiam2023gpt} API on a subset of test sets,
achieving only less than $5\%$ F1 score.
This performance disparity likely 
stems from the challenges that language models face when generating accurate lengthy texts directly for attributes such as education/work experiences. 
However, with the highest F1 score in Table \ref{tb:profiling-long} being 45.09\%, there is still room for improvement in extraction performance. 
Exploring the fusion of large language models (LLMs) and span-based methods presents a promising research avenue.

%% file: chapter/entity-tagging.tex
\subsection{Entity Tagging}

For relation enrichment, this subsection presents the results of scholar interest extraction.

\vpara{Baselines.}
For scholar interest extraction, 
we employ competition-winning solutions 
and methods relying on pre-trained models.
These approaches follow a common principle: 
they gauge the similarity between authors in the test and training sets, 
using the weighted interest tags of training authors for the authors in the test set. 
The variations among baselines lie in how they compute author similarity. 
(1) \textbf{LSI}~\cite{deerwester1990indexing} employs bag of words and TF-IDF for paper texts,
reducing dimensions with the LSI model.
(2) \textbf{ACA}\footnote{https://github.com/geekinglcq/aca}:
utilizes more paper attributes, including titles, citations, and venues,
for a more nuanced author similarity calculation. 
(3) pre-training models: 
leverage models like 
\textbf{Sentence-BERT (S-BERT)}~\cite{reimers2019sentence},
\textbf{SimCSE}~\cite{gao2021simcse}, \textbf{E5}~\cite{wang2022text},
\textbf{BERT}~\cite{devlin2019bert}, \textbf{GTE}~\cite{li2023towards},
\textbf{BGE}\footnote{\url{https://github.com/FlagOpen/FlagEmbedding}},
and \textbf{Sentence-T5 (S-T5)}~\cite{ni2022sentence}
to encode paper texts for similarity measurement.

\vpara{Evaluation Metrics.}
For scholar interest extraction, we calculate the overlap ratio 
between predicted and ground-truth tags. 

\beq{\nonumber
    \text{Accuracy} = \frac{1}{N} 
    \sum_{i=1}^{N} \frac{|T_i \cap T_i^{*}    |}{
        |T_i^{*}|
    }
}

\noindent where $N$ is the number of scholars,
$T_i^{*}$ is the annotated interest set of the $i$-th scholar, 
and $T_i$ is the predicted interest set of the $i$-th scholar. 
We pick the $3$ closest tags to the author for evaluation.
For paper topic classification, 
we measure multi-classification accuracy.

\begin{figure}[t]
	\centering
	\includegraphics[width=6.5cm]{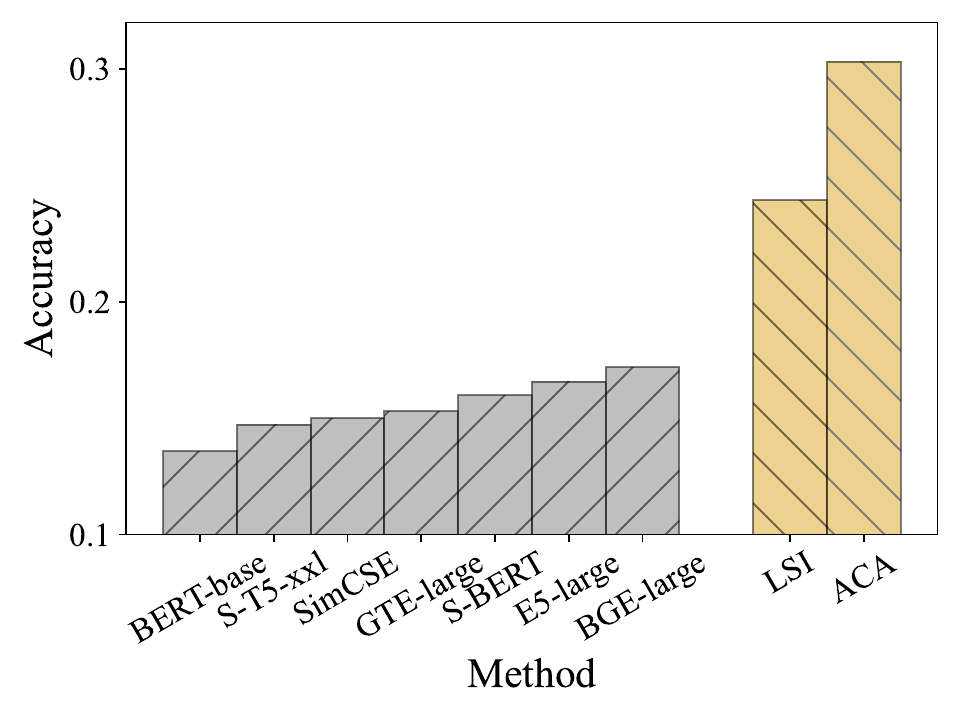}	
	\caption{
        Results of scholar interest extraction.
        \textmd{SimCSE: SimCSE-RoBERTa-large, S-BERT: all-MiniLM-L6-v2.}
    }
	\label{fig:author-tagging-results}
\end{figure}

\hide{

\begin{table}[t]
    \centering
    \caption{Results of paper topic classification ($\%$)}
    \begin{tabular}{ccc}
        \toprule
        Method & Test Acc. & Valid. Acc. \\
        \midrule
        SGC & 34.08 & 31.44 \\
        SIGN & 26.25 & 24.99 \\
        GraphSAGE & \textbf{59.57} & \textbf{57.12} \\
        \bottomrule
    \end{tabular}
    \label{tb:paper_tagging}
\end{table}

}

\vpara{Experimental Results.}
Figure \ref{fig:author-tagging-results} presents the results of scholar interest extraction.
Initial attempts to classify scholars' paper texts using research interest tags as labels yielded unsatisfactory results, 
likely due to the large number of tags and limited training data.
The methods compared in Figure \ref{fig:author-tagging-results} rely on author similarity to calculate interest tags, 
which proves more effective than text classification. 
Notably, encoding with pre-trained models directly is less effective than LSI, 
highlighting the effectiveness of shallow semantic models.
Additionally, models focused on sentence embedding outperform general pre-trained models like BERT.
The ACA method, which leverages various author attributes such as venues and citing papers, 
yields improved prediction results. 
However, the overall accuracy remains low, 
indicating a challenge in accurate classification with a large number of interest tags.

\hide{
For paper topic classification, GraphSAGE significantly outperforms SGC and SIGN. 
The reason may be that GraphSAGE has more parameters and stronger expressive ability, 
and the neighbor sampling strategy may improve its generalization ability. 
However, the training time of GraphSAGE is also significantly longer than that of SGC and SIGN. 
Therefore, how to balance efficiency and effectiveness on large-scale graph data is still a problem worthy of research. 
The classification effect of SGC is better than that of SIGN. 
The reason may be that the graph convolution filter designed by SIGN is not suitable for paper classification tasks. 
The simple convolution scheme of SGC can already capture the topic information of papers to a large extent.
}

%% file: chapter/academic-recommendation.tex
\subsection{Academic Recommendation}

For academic knowledge acquisition,
this subsection presents the results of paper recommendation and reviewer recommendation.

\vpara{Baselines.}
We compare various recommendation algorithms:
(1) \textbf{TF-IDF}, (2) Toronto paper matching system (\textbf{TPMS})~\cite{charlin2013toronto} for reviewer recommendation task,
(3) Variational autoencoder (VAE)-based item-based collaborative filtering method \textbf{Mult-VAE}~\cite{liang2018variational},
(4) Graph filtered-based collaborative filtering method \textbf{GF-CF}~\cite{shen2021powerful},
and (5) Graph neural networks (GNN)-based collaborative filtering methods \textbf{NGCF}~\cite{wang2019neural}
and \textbf{LightGCN}~\cite{he2020lightgcn}.

\vpara{Evaluation Metrics.}
Like the standard recommendation task~\cite{zhang2023apegnn, zhang2024recdcl}, 
we adopt Recall@20 and NDCG@20 as evaluation metrics for paper/reviewer recommendation.

\begin{table}[t]
    \centering
    \caption{Performance of paper recommendation.}
    \begin{tabular}{ccc}
        \toprule
        Method & Recall@20 & NDCG@20 \\
        \midrule
        Mult-VAE & 0.1088 & 0.0282 \\
        GF-CF & \textbf{0.2067} & \textbf{0.1044} \\
        NGCF & 0.1651 & 0.0823 \\
        LightGCN & 0.1950 & 0.0985\\
        \bottomrule
    \end{tabular}
    \label{tb:exp:paper_rec}
\end{table}

\vpara{Experimental Results.}
Table \ref{tb:exp:paper_rec} shows the paper recommendation performance. 
GF-CF outperforms other methods, 
highlighting the effectiveness of graph filters. 
GNN-based methods exceed Mult-VAE, 
demonstrating the value of high-order graph structures. 
LightGCN performs better than NGCF, 
which confirms the redundancy of some GNN modules in NGCF. 
However, 
there is room for improvement in recommendation accuracy.
How to use the attributes of papers and users to capture the dynamic interest changes of users is a difficult point.

\begin{table}
    \centering
    \caption{Performance of reviewer recommendation.}
    \begin{tabular}{ccc}
        \toprule
        Method & Recall@20 & NDCG@20 \\
        \midrule
        TF-IDF & 0.0016 & 0.0001 \\
        TPMS & 0.0220 & 0.008 \\
        GF-CF & \textbf{0.0382} & 0.0203 \\
        LightGCN & 0.0371 & \textbf{0.0234} \\
        \bottomrule
    \end{tabular}
    \label{tb:exp:reviewer_rec}
\end{table}

Table \ref{tb:exp:reviewer_rec}
reports the reviewer recommendation performance. 
GF-CF and LightGCN outperform TF-IDF, 
underscoring the importance of leveraging graph structures. 
The performance of TPMS is unsatisfactory because TPMS is also mainly based on TF-IDF text similarity.
However, most methods fall short of 
fully utilizing the multidimensional attributes of papers and reviewers' research interests. 
This highlights the need for further research in the reviewer recommendation task.

%% file: chapter/academic-qa.tex
\subsection{Academic Question Answering}

For academic knowledge acquisition, this subsection introduces the results of academic question answering.

\vpara{Baselines.}
We adopt sparse and dense retrieval methods:
(1) Sparse retrieval methods: \textbf{BM25},
(2) Dense retrieval methods: 
\textbf{DPR-FT} (full fine-tuning of Dense Passage Retriever (DPR)~\cite{karpukhin2020dense}),
\textbf{DPR-PT2} (parameter-efficient fine-tuning of DPR with P-Tuning v2~\cite{liu2022ptuning}),
\textbf{ColBERT-FT} (full fine-tuning of ColBERT~\cite{khattab2020colbert}),
\textbf{ColBERT-PT2} (parameter-efficient fine-tuning of ColBERT with P-Tuning v2), and
\textbf{LLM-Embedder}~\cite{zhang2023retrieve} (a fine-tuned LLM based on various retrieval-related tasks).

\vpara{Evaluation Metrics.}
Hit@K is used to measure retrieval accuracy, 
reporting if the top $K$ retrieved papers contain the correct answer. 
The average Hit@K across all questions is reported.

\begin{figure}[t]
	\centering
	\includegraphics[width=6cm]{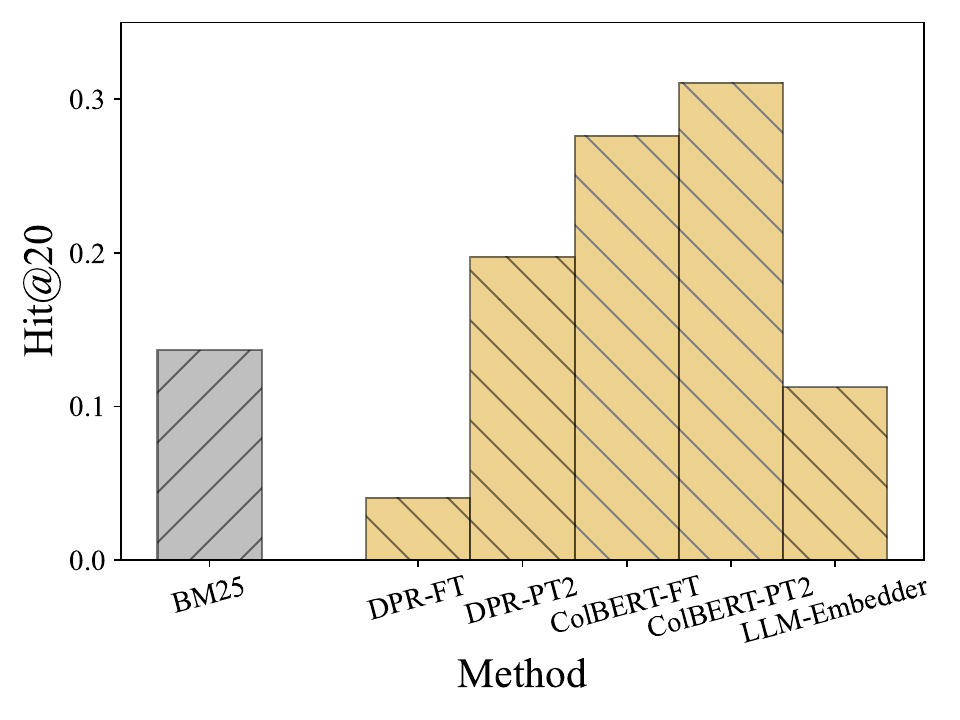}	
	\caption{
        Results of academic question answering.
    }
	\label{fig:academic-qa}
\end{figure}

\vpara{Experimental Results.}
Figure \ref{fig:academic-qa} presents the results of the OAG-QA dataset. 
Generally speaking, dense retrieval methods outperform sparse retrieval methods.
ColBERT-based methods are significantly better than DPR-based ones. 
This shows that by employing late interaction patterns and multi-vector representations, 
ColBERT models the correlation between questions and papers better. 
Interestingly, efficient parameter fine-tuning methods excel over full fine-tuning,
possibly due to better knowledge retention and generalization from the pre-trained model.
The effect of LLM-Embedder suggests there still exists noticeable gap between LLM and academic retrieval.
Overall, these methods' retrieval effects are suboptimal, 
suggesting room for improvement.

%% file: chapter/paper-source-tracing.tex
\subsection{Paper Source Tracing}

For academic source tracing, this subsection presents the results of paper source tracing.

\vpara{Baselines.}
We compare three types of methods.
(1) Statistical methods:
\textbf{Rule} (employing regular expressions to extract references 
appearing near signal words like ``motivated by'' or ``inspired by''),
and Random Forest (\textbf{RF}) (following~\cite{valenzuela2015identifying}, extracting statistical features about citations, citing positions, text similarity, etc.,
and using RF to predict the importance of references).
(2) Graph-based methods:
\textbf{LINE}~\cite{tang2015line} and \textbf{NetSMF}~\cite{qiu2019netsmf} train paper embeddings in citation networks 
and then calculates the cosine similarity between the paper embedding and the reference embedding to measure the importance of references.
(3) Pre-training methods:
extract the contextual text where each reference appears in the full texts, 
encode the text with the pre-training models, 
and use the reference annotation results in the training set 
for fine-tuning. 
The pre-training models considered include  \textbf{BERT}~\cite{devlin2019bert},
\textbf{SciBERT}~\cite{beltagy2019scibert}, \textbf{Galactica-standard}~\cite{taylor2022galactica}, and \textbf{GLM}~\cite{du2022glm}.
We also adopt three SOTA closed-source models: \textbf{GPT-3.5}~\cite{openai2022chatgpt}, \textbf{GPT-4}~\cite{achiam2023gpt}, and \textbf{Claude-instant}~\cite{anthropic2023claude}.
For both open-source and closed-source LLMs, we input the context of a referenced paper and query the model to assess the reference's significance. For instance, we ask, ``Given the context ..., is the current reference important?'' Closed-source LLMs perform this task using zero-shot evaluation.

\vpara{Evaluation Metrics.}
A paper may have one or more \textit{ref-sources}. 
For each reference of the paper $p$, 
an importance score between $[0, 1]$ needs to be output.
For each paper $p$ to be traced, 
its reference list is encoded as 0-1 based on the labeling results 
(1 if it's \textit{ref-source}, 0 otherwise).
By comparing the prediction result of each reference with its labeling result, 
we compute the Mean Average Precision (MAP). 
The average MAP across different papers serves as the evaluation metric.

\begin{figure}[t]
	\centering
	\includegraphics[width=6cm]{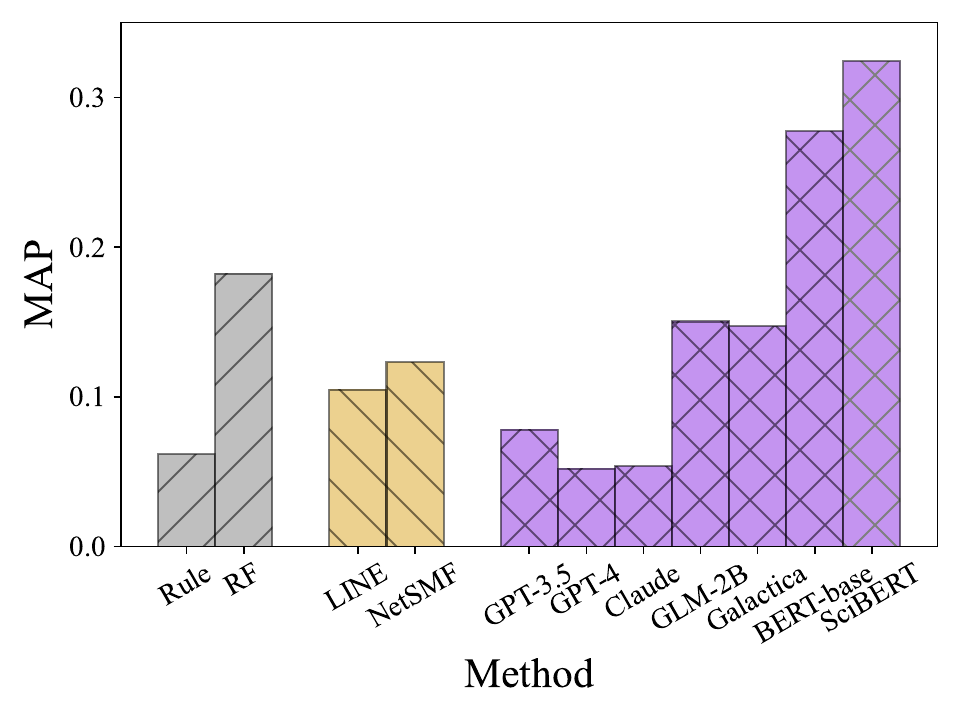}	
	\caption{
        Results of paper source tracing.
    }
	\label{fig:pst_results}
\end{figure}

\hide{
\begin{table}[t]
    \centering
    \caption{Results of paper source tracing}
    \begin{tabular}{p{1.7cm}<{\centering} p{1.7cm}<{\centering}}
        \toprule
        Method & MAP \\
        \midrule
        Rule & 0.0565 \\
        ProNE & 0.1289 \\
        BERT & \textbf{0.1294} \\
        \bottomrule
    \end{tabular}
    \label{tb:exp:paper_source_trace}
\end{table}
}

\vpara{Experimental Results.}
Figure \ref{fig:pst_results} presents the results of paper source tracing. 
Among all methods, SciBERT delivers the best performance, 
indicating the efficacy of pre-trained language models.
RF outperforms the Rule method,
underscoring the effectiveness of feature engineering.
Graph-based methods achieve average performance,
possibly owing to the ignorance of the contextual information of references.
The Rule-based approach's performance is subpar, 
likely due to many important references lacking surrounding signal words like ``inspired by'', 
resulting in a low recall.
Surprisingly, finetuned SciBERT and BERT-base outperform larger models like GLM-2B, Galactica-standard,
and closed-source LLMs.
The reason may lie in two aspects.
First, the training objective of the mask language model is more suitable for this context understanding task.
Second, API-based models may not be well-trained on similar tasks.
A potential future direction could involve merging graph-based and text-based methods for paper source tracing.
Note that the current methods' results are not yet satisfactory, 
indicating ample room for further exploration. 

%% file: chapter/academic-influence-prediction.tex
\subsection{Paper Influence Prediction}

For academic influence prediction, this subsection presents the results of paper influence prediction.

\vpara{Baselines.}
We select the following methods:
(1) \textbf{Citation}: is based on the paper citation number of known years;
(2) \textbf{Random Forest (RF)}~\cite{breiman2001random}:
defines features as the paper citation number per year and the total number of citations;
(3) \textbf{GBDT}~\cite{friedman2001greedy}: uses the same features as RF;
(4) \textbf{PageRank}~\cite{page1999pagerank}: calculates papers' PageRank score based on paper citation networks;
(5) \textbf{GraphSAGE}~\cite{hamilton2017inductive}:
performs semi-supervised classification on the paper citation network. 
Additionally, we consider graph-based node importance prediction methods: 
(6) \textbf{GENI}~\cite{park2019estimating} and
(7) \textbf{RGTN}~\cite{huang2021representation}.

\vpara{Evaluation Metrics.}
We predict for each venue to determine whether a paper would be awarded,
with labels being $0$ or $1$
indicating whether the paper is awarded or not.
Mean Average Precision (MAP) is calculated by 
comparing the predicted probability of winning the award with the ground truth label, 
and the mean MAP across different venues is used as the evaluation metric.

\begin{table}[t]
    \centering
    \caption{Results of paper influence prediction.}
    \begin{tabular}{p{1.7cm}<{\centering} p{1.7cm}<{\centering}}
        \toprule
        Method & MAP \\
        \midrule
        Citation & 0.6413 \\
        RF & 0.5409 \\
        GBDT & 0.5725 \\
        PageRank & \textbf{0.6504} \\
        GraphSAGE & 0.0811 \\
        GENI & 0.1262 \\
        RGTN & 0.0279 \\
        \bottomrule
    \end{tabular}
    \label{tb:exp:tot_pred_results}
\end{table}

\vpara{Experimental Results.}
Table \ref{tb:exp:tot_pred_results}
presents results for paper influence prediction.
Table \ref{tb:exp:tot_pred_results} shows PageRank performing best, as it considers the influence of citing papers, 
unlike the citation method that treats each citing paper equally. 
Traditional classifiers (RF and GBDT) are inferior to methods using only total citations,
indicating that total citations are a very important indicator. 
The features added by the classifier may dilute the effect of total citations. 
GraphSAGE's poor performance may be 
due to its inability to capture paper-influence factors like citation count.
GENI outperformed GraphSAGE, but both methods were less effective than Citation and PageRank methods. 
This could be due to their implicit incorporation of citation statistics and the severe class imbalance problem (positive vs. negative $< 1:100$).
Thus, identifying factors beyond citation count remains a challenge in predicting papers' breakthrough innovation.

%% file: leaderboard.tex
\section{\contestname}

To promote the engagement of the research community and the development of \benchname,
we also introduce
the Open Academic Data Challenge (\contestname) and 
set up a regular leaderboard for up-to-date \benchname\footnote{\url{https://www.biendata.xyz/kdd2024/}}.
\contestname
currently contains three challenging academic tasks:
incorrect assignment detection for author name disambiguation (IND),
academic question answering (OAG-AQA), and paper source tracing (PST).

Specifically, 
given the paper assignments of each author and paper metadata, 
IND aims to 
detect paper assignment errors for each author.
Given professional questions and a pool of candidate papers, 
OAG-AQA hopes to 
retrieve the most relevant papers to answer these questions.
As mentioned earlier, 
given the full texts of each paper, 
PST aims to 
automatically trace the most significant references that have inspired a given paper.

\contestname was deployed at KDD Cup 2024 and attracted more
than 800 team registrations globally.
Following the previous successful conventions,
submissions are required to provide 
source codes, technical reports, and contact information
for better knowledge sharing and iteration.
We are periodically updating the datasets,
including annotating new assignment errors, 
crawling new academic question and answer pairs,
and collecting new reading records for PST.

%% file: conclusion.tex
\section{Conclusion}

The attention of the research community to academic benchmarks remains limited,
even if academic tasks offer various challenges and applications of immense impact.
Thus, this paper introduces \benchname
to carefully annotate large-scale OAG for the full life cycle of academic graph mining.
\benchname now includes 10 tasks, \datasetnum datasets, 
\methodnum baseline models, and \expnum experimental results.
In the future, we plan to continually maintain and enhance \benchname
by updating up-to-date datasets regularly from real scenarios,
adding more practical tasks, and exploring interactive evaluation metrics.
\benchname is always open for contributions from communities
by adding new tasks or datasets, developing cutting-edge algorithms or foundation models for various tasks, etc.
\hide{
We plan to continually maintain and enhance the OAG-Bench in the following ways.
First, we will update up-to-date datasets regularly from real scenarios,
such as expanding OAG-QA by crawling QA pairs continuously 
and checking new name disambiguation results by professional annotators.
Second, we will add more practical tasks.
For example, more generation-based tasks can be incorporated, such as literature review generation and paper summarization.
Third, we will explore appropriate evaluation metrics. 
How to minimize the time cost of human feedback and how to evaluate interactively is worth studying.
}

%% file: appendix.tex
\section{Results of Additional Tasks}
\label{app:other-tasks}

\subsection{Entity Alignment}
\label{app:author_alignment}

We provide the experimental results of venue alignment, 
affiliation alignment, and author alignment in this subsection.

\vpara{Baselines.}
Venue alignment and affliation alignment are short text matching tasks.
We compare different types of matching methods:
(1) traditional machine learning methods (\textbf{SVM})
using Jaccard index and TF-IDF similarity as input features;
(2) shallow neural network-based matching methods
including CNN-based matching model (\textbf{\modeloagcnn})~\cite{zhang2019oag}
and RNN-based matching model (\textbf{\modeloaglstm})~\cite{zhang2019oag};
(3) matching models based on pre-trained models (\textbf{Ditto}~\cite{li2020deep}
with pre-trained models BERT~\cite{devlin2019bert}, ALBERT~\cite{lan2019albert}, 
RoBERTa~\cite{liu2019roberta}, DeBERTa~\cite{he2022debertav3}, 
LaBSE~\cite{feng2022language}, and GLM~\cite{du2022glm}).

Author alignment can further take authors' structural information into account.
Apart from SVM, \modeloagcnn, and \modeloaglstm,
we additionally select \modeloaggat~\cite{zhang2019oag} model for author alignment,
which constructs a subgraph for each candidate pair,
and then uses heterogeneous graph attention networks to learn hidden representations for classification.
For each candidate author pair, we leverage published papers and venues of authors as features.

\vpara{Evaluation Metrics.}
Entity alignment is a binary classification problem.
We take F1 and AUC as the evaluation metric.

\begin{table}[t]
    \newcolumntype{C}{>{\centering\arraybackslash}p{2.2em}}
    \caption{
        \label{tb:entity-matching} Alignment results of different types of entities ($\%$).
    }
    \centering
    \renewcommand\arraystretch{1.0}
    \begin{tabular}{c|@{~ }*{1}{CC|}*{1}CC}
        \toprule[1.2pt]
        &\multicolumn{2}{c|}{Venue}
        &\multicolumn{2}{c}{Affiliation}
    \\
    \cmidrule{2-3} \cmidrule{4-5} 
    {Method}   & {F1} & {AUC} & {F1}  & {AUC}   \\
    \midrule
    SVM & 82.63 & 91.87 & 68.47 & 70.93 \\
    \modeloagcnn & 83.46 & 94.07 & 69.06 & 69.12 \\
    \modeloaglstm & 85.03 & 95.33 & 67.76 & 72.37 \\
    Ditto-BERT-base & 89.33 & 95.00 & 70.38 & 78.65 \\
    Ditto-ALBERT-base & 88.05 & 96.06 & 61.90 & 70.96 \\
    Ditto-RoBERTa-large & 89.47 & 97.22 & 71.98 & 79.02 \\
    Ditto-DeBERTa-base & \textbf{94.27} & 96.73 & 72.78 & 80.37\\
    Ditto-DeBERTa-large & 89.44 & \textbf{97.94} & \textbf{82.18} & \textbf{89.76} \\
    Ditto-LaBSE & 90.79 & 96.93 & 71.06 & 78.55 \\
    Ditto-GLM-RoBERTa & 78.82 & 93.04 & 61.11 & 67.93 \\
    \bottomrule[1.2pt]
\end{tabular}
\end{table}

\begin{table}[t]
    \centering
    \caption{Performance of author alignment ($\%$).}
    \begin{tabular}{ccc}
        \toprule
        Method & F1 & AUC \\
        \midrule
        SVM & \textbf{90.14} & 93.67 \\
        \modeloagcnn & 67.72 & 66.78\\
        \modeloaglstm &  74.80 & 77.58 \\
        \modeloaggat & 89.62 & \textbf{97.32} \\
        \bottomrule
    \end{tabular}
    \label{tb:exp:author_alignment}
\end{table}

\vpara{Experimental Results.}
Table \ref{tb:entity-matching} presents results for various matching models
for venue alignment and affiliation alignment.
The methods for author alignment utilize more structural information, 
with corresponding results presented in Section \ref{app:author_alignment}.
Venue/affiliation alignment is essentially a short text matching task.
SVM performs slightly poorer than \modeloagcnn and \modeloaglstm,
possibly due to the inability of SVM features to capture word order information.
Conversely, \modeloagcnn and \modeloaglstm can capture the contextual dependence of word sequence.
Overall, the top-performing models on both datasets utilize pre-training, 
indicating the promising potential of pre-trained language models for entity alignment.
Among all methods that employ pre-trained models, 
DeBERTa-large delivers the best performance on two tasks, particularly on affiliation alignment,
indicating that DeBERTa-large effectively encodes additional semantic knowledge beyond affiliations' surface names.
Note that the trends of AUC and F1 in the table are sometimes inconsistent.
Given that F1 requires a threshold setting, 
AUC is a more reliable metric when discrepancies arise between the two.

The results of author alignment are shown in Table \ref{tb:exp:author_alignment}. 
We observe that methods using authors' structure information (SVM and \modeloaggat)
are significantly better than methods not using structure information (\modeloagcnn and \modeloaglstm).
In the future, more author pairs with ambiguous names will be added
to increase the difficulty of this task.

\subsection{Author Name Disambiguation}

We present the detailed task description and experiments for author name disambiguation in this subsection.
Two subtasks of author name disambiguation are defined as follows.

\theoremstyle{problem}
\begin{problem}{\bf From-scratch Name Disambiguation (SND).} 
Given a collection of papers associated with identically-named authors,
the goal is to cluster these papers into distinct groups,
where each group should represent papers by the same author, 
while different groups signify papers by different authors.
\end{problem}

\theoremstyle{problem}
\begin{problem}{\bf Real-time Name Disambiguation (RND).} 
Given a collection of unassigned papers and 
a set of authors (each author includes attributes such as affiliations, research interests, published papers, etc.),
the goal is to assign these papers to the correct author or return empty
(meaning that no author can be matched).
\end{problem}

\vpara{Datasets.}
The WhoIsWho dataset includes \num{72609} authors with \num{2459} names,
and \num{1102249} associated papers. 
The authorship between papers and authors is manually annotated.
For the two subtasks, the training set contains the mapping relationship among the author name --- author ID --- paper ID, and the metadata of the paper (such as title, author name, published venue, etc.).
For the SND task, the validation set and test set contain the name to be disambiguated 
and the papers associated with the name, 
and the goal is to cluster the papers into different groups.
For the RND task, the validation and test sets involve unassigned papers, 
and the goal is to assign papers to existing authors or return NIL.

\vpara{Baselines.}
We select the winning solutions of the latest competition\footnote{\url{https://www.biendata.xyz/competition/whoiswho1/}, \url{https://www.biendata.xyz/competition/whoiswho2/}} for comparison.
Specifically, the compared SND methods include
\textbf{ECNU\_AIDA}, \textbf{Complex808}, and \textbf{liub}.
The three methods follow a similar framework.
They first encode the semantic features of papers via pre-trained models.
Then, they construct a heterogeneous network according to the heterogeneous attributes of papers.
Next, they use random walks based on meta-paths to generate the structural representation of papers.
The semantic representation and structural representation can separately generate two paper similarity matrices.
Finally, they use the DBSCAN clustering algorithm to cluster the papers to obtain the clustering result.
The difference between the three methods lies in:
(1) \textbf{ECNU\_AIDA} and \textbf{Complex808} use pre-trained Word2Vec model~\cite{mikolov2013distributed} to obtain the semantic representation of papers, 
while \textbf{liub} uses OAG-BERT~\cite{liu2022oag} to obtain the semantic representation of papers;
(2) \textbf{ECNU\_AIDA} and \textbf{liub} use co-author and co-organization relations between two papers for random walks in heterogeneous networks, 
while \textbf{Complex808} additionally introduces co-venue and co-keyword relations between two papers with a certain probability for random walks.

\hide{
\begin{itemize}[leftmargin=*]
    \item \textbf{ECNU\_AIDA:}
    This method uses pre-trained Word2Vec model~\cite{mikolov2013distributed} to obtain the semantic representation of papers, 
    and then constructs a heterogeneous network according to the heterogeneous attributes of papers, 
    and uses random walks based on meta-paths to generate the structural representation of papers. 
    The semantic representation and structural representation can separately generate two paper similarity matrices.
    The two similarity matrices are added to use the DBSCAN clustering algorithm to cluster the papers to obtain the clustering result.
    \item \textbf{Complex808:}
    This method is similar to the ECNU\_AIDA framework.
    For random walks in heterogeneous networks, 
    this method uses co-author and co-organization relations between two papers. 
    The method also finds that if two papers are published in the same venue, or have the same keywords, 
    since these types of relationships may introduce noise, 
    the method selects these two types of meta-paths with a certain probability for random walk.
    \item \textbf{liub:} 
    This method is similar to the ECNU\_AIDA framework.
    The difference is that the semantic features of this method are generated using OAG-BERT [186]. 
    This method does not take advantage of the relationship of papers published in the same venue to generate heterogeneous networks.
\end{itemize}
}

Compared RND methods include:
(1) \textbf{kingsundad}:
employs three similarity features: handcrafted, OAG-BERT, and RBF-kernel interaction matching~\cite{xiong2017end}. 
These features are input into multiple classifiers like XGBoost~\cite{chen2016xgboost}, 
LightGBM~\cite{ke2017lightgbm}, and CatBoost~\cite{dorogush2018catboost} for ensemble learning.
(2) \textbf{AlexNE}:
introduces various name encoding techniques, like abbreviated encoding, to enhance recall. 
Unlike the Kingsundad method, it uses both OAG-BERT and GloVe~\cite{pennington2014glove} to generate semantic paper representations. 
It constructs a large graph connecting unassigned papers to existing ones via keywords 
and generates node representations using Node2Vec~\cite{grover2016node2vec}. 
(3) \textbf{Data Magician}:
is a feature engineering-based method 
that uses features from keywords, affiliations, co-authors, years, and more. 
Notably, it defines a time-weighted paper similarity method to account for changing research interests.

\vpara{Evaluation Metrics.}
For the SND task and for each name,
we evaluate the widely-adopted pairwise-F1.

\begin{align}\nonumber
    \begin{split}
            PairwisePrecision=\frac{
        \#PairsCorrectlyPredictedToSameAuthor
    }{
        \#TotalPairsPredictedToSameAuthor
    }\\
    PairwiseRecall=\frac{
        \#PairsCorrectlyPredictedToSameAuthor
    }{
        \#TotalPairsToSameAuthor
    }\\
    PairwiseF1 = \frac{
        2 \times PairwisePrecision \times PairwiseRecall
    }{
        PairwisePrecision + PairwiseRecall
    }
\end{split}
\end{align}

\noindent The overall pairwise-F1 is the mean of pairwise-F1 of all names.

For the RND task, we first calculate precision and recall for each author's unassigned papers,
and then take the weighted average of precision and recall of different authors
according to the paper count of each author
to compute the overall F1.

\begin{table}[t]
    \centering
    \newcolumntype{C}{>{\centering\arraybackslash}p{2cm}}
    \caption{Disambiguation results on WhoIsWho dataset ($\%$).}
    \subcaptionbox{Results of SND methods.
    \label{tb:whoiswho_scratch}}{
        \begin{tabular}{C p{1cm} <{\centering}}
            \toprule
            Method & F1 \\
            \midrule
            ECNU\_AIDA & \textbf{89.140} \\
            Complex808 & 88.594 \\
            liub & 88.580\\
            \bottomrule
        \end{tabular}
    }
    \subcaptionbox{Results of SND methods.
    \label{tb:whoiswho_online}}{
        \begin{tabular}{C p{1cm} <{\centering}}
            \toprule
            Method & F1\\
            \midrule
            kingsundad & \textbf{93.492} \\
            AlexNE & 93.136 \\
            Data Magician & 92.850 \\
            \bottomrule
        \end{tabular}
    }
    \label{tb:whoiswho_exp}
\end{table}

\vpara{Experimental Results.}
Table \ref{tb:whoiswho_scratch} and Table \ref{tb:whoiswho_online} report
the performance of SND and RND methods, respectively.
For SND methods, 
despite using OAG-BERT for semantic paper representation, 
the liub method underperforms, 
suggesting the need for further exploration of applying large language models. 
In addition, 
given the similarity of the three methods' frameworks, 
how to break away from paper representations based on semantic and structural dimensions to calculate paper similarity 
and then use the DBSCAN algorithm to cluster papers is also worthy of further study.

For RND methods, 
all three methods yield good results. 
AlexNE attempts to introduce a graph structure to help disambiguate unassigned papers, 
which is a less explored direction. 
Data Magician uses time-based paper similarity features for complex name disambiguation scenarios. 
Further research is needed for challenging situations like co-authors with identical names or affiliation shifts, 
and for designing a robust paper-author matching model, 
considering potential incorrect assignments in existing authors' papers.

\subsection{Scholar Profiling}
This subsection presents the task description and experiments for 
search engine-based scholar profiling.

\theoremstyle{problem}
\begin{problem}{\bf Search Engine-based Scholar Profiling.} 
    Given a scholar's name, affiliation, 
    and one's search engine records 
    (using ``name + affiliation'' to query
    and extracting up to $2$ search pages and up to $20$ snippets),
    the goal is to extract the portrait information of the scholar,
    including homepage, gender, and position.
\end{problem}

\vpara{Datasets.}
\textbf{CCKS2021-En}\footnote{\url{https://www.biendata.xyz/competition/ccks_aminer_profiling/}}:
This dataset is 
an English subset of the CCKS 2021 scholar profiling track from AMiner.
It contains $9{\small,}221$ scholar portraits,
randomly divided into
$5{\small,}557$ for training,
$1{\small,}833$ for validation,
and $1{\small,}831$ for testing.

\vpara{Baselines.}
Drawing from the winning solutions of the CCKS 2021
and recent named entity recognition (NER) methods,
we select the following baselines for search engine-based scholar profiling:
(1) \textbf{SML}: employ manual features and traditional classifiers.
Specifically, for gender prediction, 
\textbf{SML-esb} extracts features such as
the frequency of ``his'' and ``her''
and uses various classifiers for voting.
For homepage extraction,
\textbf{Logistic Regression (LR)} and \textbf{XGBoost}
extract features like the appearance of signal words (such as ``edu'' and ``academic'') for classification.
(2) \textbf{Rule}:
utilizes regular expressions
and voting for position extraction.
(3) \textbf{BI-LSTM-CRF}~\cite{huang2015bidirectional}:
uses a BI-LSTM layer 
and a CRF layer for sequence labeling in position extraction.
(4) Pre-training methods: 
We design different inputs for pre-training models for each attribute.
For gender prediction and position extraction, the scholar's name, affiliation, and webpage texts are concatenated as inputs.
For homepage extraction, the scholar's name and the candidate URL are concatenated as inputs.
We fine-tune pre-trained models for classification,
including \textbf{BERT}~\cite{devlin2019bert},
\textbf{RoBERTa}~\cite{liu2019roberta},
\textbf{DeBERTa}~\cite{he2022debertav3},
\textbf{ALBERT}~\cite{lan2019albert},
\textbf{ChatGLM}\footnote{\url{https://github.com/THUDM/ChatGLM-6B}},
\textbf{LLaMA}~\cite{touvron2023llama}.

\vpara{Evaluation Metrics.}
Following the competition,
accuracy is used to measure exact matches between predictions and ground truths.

\begin{table}[t]
    \centering
    \caption{
        Results of search engine-based scholar profiling ($\%$).
    }
    \label{tb:profiling-basic}
    \begin{tabular}{cccc}
        \toprule
        Method & Gender & Homepage & Position \\
        \midrule
        SML-esb & 71.67 & -- & -- \\
        LR & -- & 19.37 & --\\
        XGBoost & -- & 20.8 & --\\
        Rule & -- & -- & 71.6 \\
        BI-LSTM-CRF & -- & -- & 85.10\\
        \midrule
        BERT-base & 96.12 & 18.91 & 83.23\\
        RoBERTa-base & 96.40 & 20.65 & 83.78 \\
        DeBERTa-base & 96.50 & 21.11 & \textbf{85.14} \\
        DeBERTa-v3-large & \textbf{96.56} & 18.21 & 79.34 \\
        ALBERT-base & 95.85 & 16.38 & 84.33 \\
        ChatGLM-6B-LoRA & 96.40 & 26.83 & 78.15 \\
        LLaMA-7B-LoRA & 70.07 & \textbf{26.93} & 79.14 \\
        \bottomrule
    \end{tabular}
\end{table}

\vpara{Experimental Results.}
Table \ref{tb:profiling-basic}
displays extraction results for search engine-based scholar profiling.
In Table \ref{tb:profiling-basic}, 
pre-trained models outperform traditional methods,
showcasing the expressive capacity and 
effectiveness of pre-trained models without manual feature design. 
BI-LSTM-CRF and partial pre-trained models exhibit similar performance in position extraction,
showing the suitability of both sequence labeling and pre-trained models. 
Large generative models excel in homepage extraction, 
though accuracy remains modest.

\subsection{Entity Tagging}
\label{app:paper_tagging}

We provide the experimental setup and results of the paper topic classification as follows.

\vpara{Baselines.}
We adopt three GNN methods as baselines,
including \textbf{SGC}~\cite{wu2019simplifying}, \textbf{SIGN}~\cite{rossi2020sign},
and \textbf{GraphSAGE}~\cite{hamilton2017inductive}.

\vpara{Evaluation Metrics.}
We measure multi-classification accuracy for paper classification.

\begin{table}[t]
    \centering
    \caption{Results of paper topic classification ($\%$).}
    \begin{tabular}{ccc}
        \toprule
        Method & Test Acc. & Valid. Acc. \\
        \midrule
        SGC & 34.08 & 31.44 \\
        SIGN & 26.25 & 24.99 \\
        GraphSAGE & \textbf{59.57} & \textbf{57.12} \\
        \bottomrule
    \end{tabular}
    \label{tb:paper_tagging}
\end{table}

\vpara{Experimental Results.}
Table \ref{tb:paper_tagging} shows the results of paper topic classification.
GraphSAGE outperforms SGC and SIGN in paper topic classification due to more training parameters, 
expressive ability, and neighbor sampling strategy. 
However, its longer training time poses a challenge in balancing efficiency and effectiveness on large-scale graph data. 
SGC performs better than SIGN, possibly because SIGN's graph convolution filter is less suited for paper classification tasks, 
while SGC's simpler convolution scheme effectively captures paper topic information.
Current methods mainly leverage paper citation structure for paper topic classification,
yielding unsatisfactory results.
More content information could be incorporated to enhance the fine-grained paper tagging performance.

\subsection{Concept Taxonomy Completion}

In this subsection, we provide the experimental setup and results for concept taxonomy completion.

\vpara{Baselines.}
Some of the latest concept taxonomy completion methods are selected for comparison.
(1) \textbf{BiLinear}~\cite{sutskever2009modelling}:
uses a bilinear model to encode new and candidate concept representations, 
performing binary classification to ascertain if a candidate position owns the correct hypernym and hyponym of a new concept.
(2) \textbf{TaxoExpan}~\cite{shen2020taxoexpan}:
employs a position-augmented graph neural network 
to gauge the relationship between new concepts and candidate concept subgraphs, 
using contrastive learning
to bolster model robustness. 
We use SciBERT~\cite{beltagy2019scibert} to encode concepts for a fair comparison with the next method.
(3) \textbf{TaxoEnrich}~\cite{jiang2022taxoenrich}:
initially transforms the existing hyponymy relationship into natural language, 
using SciBERT to represent concepts.
It then employs LSTM to encode vertical concept relationships 
and an attention mechanism for sibling relationships. 
Finally, a matching model calculates the score between a new concept and a candidate position.

\vpara{Evaluation Metrics.}
Each new concept is matched with nodes in the existing concept hierarchy tree 
and sorted by similarity. 
Evaluation metrics include Hit@10 and Mean Reciprocal Rank (MRR), 
which is the average rank of the reciprocal of actual hypernyms.

\hide{
\begin{itemize}[leftmargin=*]
    \item \textbf{BiLinear~\cite{sutskever2009modelling}:}
    This method encodes new concept representations and candidate concept representations using a bilinear model 
    for binary classification to determine whether a candidate concept is a hypernym of a new concept.
    \item \textbf{TaxoExpan~\cite{shen2020taxoexpan}:}
    is one of the most state-of-the-art hypernymy prediction algorithms. 
    It utilizes a position-augmented graph neural network to measure the relationship between new concepts and candidate concept subgraphs, 
    and uses the InfoNCE~\cite{oord2018representation} loss function to enhance the robustness of the model.
    \item \textbf{TaxoEnrich~\cite{jiang2022taxoenrich}:}
    This method first converts the existing hyponymy relationship into natural language, 
    and then uses the pre-training model to generate the initial representation of the concept; 
    after that, it uses LSTM to encode the vertical relationship between concepts, 
    and uses the attention mechanism to encode the sibling relationship between concepts; 
    finally, a matching model between a new concept and a candidate position is used to calculate the matching score.
\end{itemize}
}

\begin{table}[t]
	\centering
	\newcolumntype{?}{!{\vrule width 1pt}}
	\caption{
        Results of concept taxonomy completion.
	}
	\label{tb:exp:taxonomy}
	\begin{tabular}{  c | p{1.2cm} <{\centering}? p{1.4cm} <{\centering}| p{1.4cm}<{\centering}| p{1.4cm}<{\centering}}
		\toprule
		\multicolumn{2}{c?}{Method} & {BiLinear} & TaxoExpan  & TaxoEnrich \\
		
        \midrule
		\multirow{2}*{MAG-full} 
	    & Hit@10  &  0 & \textbf{0.216} & 0.003 \\
		~ & MRR & 0.002 & \textbf{0.221} & 0.008 \\

		\midrule
		
		\multirow{2}*{MAG-CS} 
	    & Hit@10  & 0.022 & \textbf{0.301} & 0.132  \\
		~ & MRR & 0.059 & \textbf{0.376} & 0.204 \\
		\midrule
		
		\multirow{2}*{OAG-AI} 
	    & Hit@10 & 0.022 & \textbf{0.343} & 0.166 \\
		~ & MRR &  0.059 & \textbf{0.422} &  0.238 \\
		\bottomrule
	\end{tabular}
\end{table}

\vpara{Experimental Results.}
Table \ref{tb:exp:taxonomy} reports the performance of concept taxonomy completion. 
TaxoExpan 
outperforms other methods on three datasets, 
indicating the effectiveness of the position-augmented graph neural network. 
TaxoEnrich surpasses TaxoExpan in its paper, 
likely due to its more potent pre-trained representation. 
The BiLinear model's simplicity limits its expressive power, 
affecting prediction accuracy. 
The results of OAG-AI are obtained by making inferences using
the pre-trained model on MAG-CS,
maintaining similar trends as other datasets.
Hit@10 doesn't exceed 0.35 on all datasets, 
indicating the challenge of automatic taxonomy construction 
and the potential need for more information or powerful model architectures.

\subsection{Academic Influence Prediction}

In this subsection, we provide the experimental setup and results for author influence prediction.

\vpara{Baselines.}
\hide{
We select the following methods:
(1) \textbf{Citation}: is based on the paper citation number of known years;
(2) \textbf{Random Forest (RF)}~\cite{breiman2001random}:
defines features as the paper citation number per year and the total number of citations;
(3) \textbf{GBDT}~\cite{friedman2001greedy}: uses the same features as RF;
(4) \textbf{PageRank}~\cite{page1999pagerank}: calculates papers' PageRank score based on paper citation networks;
(5) \textbf{GraphSAGE}~\cite{hamilton2017inductive}:
performs semi-supervised classification on the paper citation network.
}
We adopt the following baselines:
(1) \textbf{ARIMA}: is a statistical model for time series forecasting;
(2) \textbf{Linear Regression}: defines a series of features of each author, 
including the author's annual citations in the past 20 years, 
the total number of citations, the total number of papers, 
the H-index\footnote{\url{https://en.wikipedia.org/wiki/H-index}}, 
and the estimated citation number of the author by using author citations and paper-author relations.
Then, we use the Linear Regression model to predict the number of citations of the author.
(3) \textbf{GBRT}~\cite{friedman2001greedy}: uses the same features as linear regression.
(4) \textbf{LSTM}~\cite{hochreiter1997long}: uses the features (\#citations and \#papers) of the author in the past 20 years as a time series
and uses LSTM for regression prediction.
(5) \textbf{EvolveGCN}~\cite{pareja2020evolvegcn}: models author influence prediction as a node regression problem on dynamic co-author graphs.

\vpara{Evaluation Metrics.}
\hide{
For the paper influence prediction, 
we predict for each venue to determine whether a paper would be awarded,
with labels being $0$ or $1$
indicating whether the paper is awarded or not.
Mean Average Precision (MAP) is calculated by 
comparing the predicted probability of winning the award with the ground truth label, 
and the mean MAP across different venues is used as the evaluation metric.
}
The root mean square error (RMSE) between the predicted cited number and the actual cited number is used as the evaluation metric.

\begin{table}[t]
    \centering
    \caption{ Performance of author influence prediction (RMSE).
        }
    \begin{tabular}{ccc}
        \toprule
        Method & AuthPred-2016 & AuthPred-2022 \\
        \midrule
        ARIMA & 1225 & 23920 \\
        Linear Regression & 562 & 22057 \\
        GBRT & \textbf{553} & \textbf{21777} \\
        LSTM & 1034 & 25409 \\
        EvolveGCN & 969 & 22841 \\
        \bottomrule
    \end{tabular}
    \label{tb:exp:author_inf_pred}
\end{table}

\vpara{Experimental Results.}
Table \ref{tb:exp:author_inf_pred}
present results for author influence prediction.
It reveals that 
the GBRT method has smaller prediction errors on both datasets, 
demonstrating its superior fitting ability over linear regression 
and the effectiveness of input features like author citations.
ARIMA's poor performance suggests that time-series-based statistical methods struggle to predict academic influence. 
In addition to being affected by past achievements, 
the future influence of scholars will also have dynamic and more complex factors. 
EvolveGCN outperforms LSTM, 
indicating co-author network dynamics contain factors related to author influence.
The larger prediction error on the AuthPred-2022 dataset 
could be due to differences in citation statistics between AMiner and Google Scholar, 
or the increased difficulty in predicting author influence in 2022 
due to the surge in paper numbers.

\section{Ethical Statement}

\benchname involves author-centric attributes.
We exclude those sensitive attributes such as email and profile photo,
making available attributes publicly accessible elsewhere.
For online publications, \benchname provides publicly available metadata
and very few parsed full-texts of open-access papers for research purposes.
For data annotation, all annotators gave their informed consent for inclusion before they participated in this study.